\documentclass[pre,superscriptaddress,showpacs,twocolumn]{revtex4}
\usepackage{graphicx}
\usepackage{graphicx,amsfonts}
\usepackage{amsmath}
\usepackage{verbatim}

\begin{document}

\newcommand{\atanh}
{\operatorname{atanh}}
\newcommand{\ArcTan}
{\operatorname{ArcTan}}
\newcommand{\ArcCoth}
{\operatorname{ArcCoth}}
\newcommand{\Erf}
{\operatorname{Erf}}
\newcommand{\Erfi}
{\operatorname{Erfi}}
\newcommand{\Ei}
{\operatorname{Ei}}

\title{The longest excursion of fractional Brownian motion : numerical evidence of non-Markovian effects}
\author{Reinaldo Garc{\'i}a-Garc{\'i}a}
\affiliation{Centro At\'omico Bariloche, 8400 S. C. de Bariloche, Argentina}
\author{Alberto Rosso}
\affiliation{Laboratoire de Physique Th\'eorique et Mod\`eles Statistiques
  (UMR du CNRS 8626), Universit\'e de Paris-Sud, 91405 Orsay Cedex, France}
\author{Gr\'egory Schehr}
\affiliation{Laboratoire de Physique Th\'eorique (UMR du
 CNRS 8627), Universit\'e de Paris-Sud, 91405 Orsay Cedex, France}

\begin{abstract}
We study, using exact numerical simulations, the statistics of the longest excursion $l_{\max}(t)$ up
to time $t$ for the fractional Brownian motion with Hurst exponent
$0<H<1$. We show that in the large $t$ limit, $\langle l_{\max}(t) \rangle \propto Q_\infty t$ where
$Q_\infty \equiv Q_\infty(H)$ depends continuously on $H$, and in a non trivial way. These
results are compared with exact analytical results obtained recently
for a renewal process with an associated persistence exponent $\theta
= 1-H$. This comparison shows that $Q_\infty(H)$ carries the clear
signature of non-Markovian effects for $H\neq 1/2$. The pre-asymptotic behavior of $\langle l_{\max}(t)\rangle$ is also discussed.
\end{abstract}

\maketitle


{\it Introduction.} In the last few years, there has been a growing interest
in the study of {\it anomalous dynamics} \cite{antoine_bouchaud, metzler_review}, where by
contrast with Brownian motion, long range temporal correlations induce
non-standard dynamical behaviors. Instead of diffusive behavior, anomalous
dynamics typically displays a non linear growth of the mean square
displacement $\langle x^{2}(t)\rangle \propto t^{2H}$, where $H \neq 1/2$ is
the Hurst exponent. Such behaviors have been observed in various experimental
situations including polymer networks \cite{leibler_polymers}, intracellular
transport \cite{caspi}, two-dimensional rotating flows \cite{weeks_fluids} or porous glasses \cite{stapf_glass}. To describe theoretically such situations, various stochastic processes have been proposed and studied. Among them, the fractional Brownian motion (fBm), initally introduced by Mandelbrot and van Ness \cite{mandelbrot}, is currently playing an increasing role in this area of research. For instance, the fBm was recently proposed to model the stochastic dynamics of a polymer passing through a pore (translocation) \cite{kardar_kantor, zoia_translocation}.

\begin{figure}[t]
\centerline{
\includegraphics[width=\columnwidth]{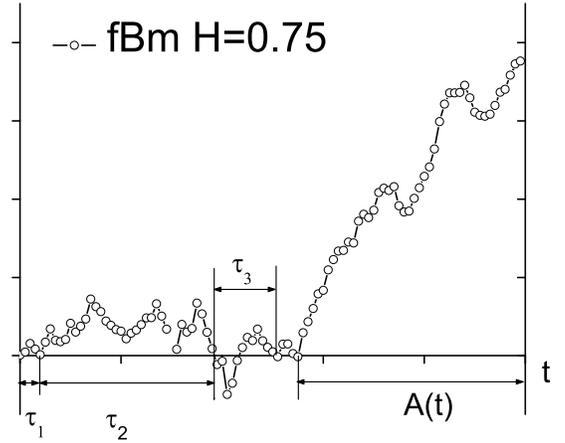}}
\caption{Intervals between zero crossings (excursions) for the fBm in the particular case $H=0.75$, which was generated numerically using the Levinson algorithm. The longest excursion $l_{\max}(t)$, studied in this paper, is defined in Eq. (\ref{def_lmax}).}
\label{Fig1}
\end{figure}

The fBm $x(t)$ is a Gaussian stochastic process characterized by the following two-time correlations
\begin{eqnarray}\label{correl}
\langle x(t_1) x(t_2) \rangle=C(t_1,t_2) = t_1^{2H} + t_2^{2H} - |t_1-t_2|^{2H} \;.
\end{eqnarray}
This implies that the incremental correlation
function is stationary, {\it i.e.} $\langle (x(t_1) - x(t_2))^2 \rangle  =
|t_1 - t_2|^{2H}$. For $H=1/2$, the process $x(t)$ is just Brownian motion (BM). For $H < 1/2$
the dynamics is subdiffusive, while it is superdiffusive for $H >1/2$. For $0 < H < 1$, fBm is a non-smooth process, {\it i.e.} it
has an infinite density of zero crossings. A relevant quantity characterizing these zero crossings is the
distribution $\rho(\tau)$ of the time intervals between consecutive zeros. In many cases, which are relevant in statistical physics, this distribution has a power law tail  $\rho(\tau) \propto \tau^{-1-\theta}$, with $\theta$ the persistence exponent \cite{godreche, satya_review}. A remarkable result for processes, Gaussian or non-Gaussian, obeying Eq. (\ref{correl}), is the {\it exact} relation $\theta = 1-H$ \cite{fbm_persistence, krug_persistence, satya_persistence}. Such processes (\ref{correl}) appear naturally in various interesting models of statistical physics. For instance, the fBm with $H=1/4$ arises as a
scaling limit of a tagged particle in a one-dimensional symmetric
exclusion process \cite{arratia}. It also describes the equilibrium temporal
fluctuations of the height field of a $d$-dimensional
Edwards-Wilkinson interface, and in that case $H = (1-d/2)/2$ 
\cite{krug_persistence}.  Another example where such a process as in Eq. (\ref{correl}), albeit non-Gaussian, appears is the Matheron-de Marsily model of hydrodynamic flows in porous media.  There it describes 
the longitudinal position of a particle in a $d+1$-dimensional layered random velocity field and in that case $H =
\max(1-d/4,1/2)$~\cite{satya_mdm}.

For $H \neq 1/2$, one expects that fBm is a non-Markov process
\cite{mandelbrot}. However the zero-crossing properties of the fBm which
have been investigated up to now have not convincingly shown the signatures of
these memory effects. For instance assuming that the intervals between
crossings are independent and identically distributed (renewal process)
 yields the correct behavior for the tail of the distribution $\rho(\tau)$  with  $\theta = 1-H$ \cite{krug_persistence_renewal}. More recently, 
on the basis of a numerical computation of the correlation function of the intervals
between successive zeros, the authors of Ref. \cite{cakir_renewal} claimed that the zero crossings properties of fBm are actually described by a renewal process, which contradicts our theoretical understanding of this process \cite{mandelbrot, krug_persistence}. One goal of the present paper is thus to exhibit a property of the fBm which instead shows that temporal correlations clearly affect the zero crossings properties of this process.

To this purpose, following a recent work 
\cite{us_prl}, we study here the statistics of the longest excursion up to
time $t$, denoted $l_{\rm max}(t)$.  
For a typical realization of the fBm $x(t)$ with
$N \equiv N(t)$ zeros in the {\em fixed} time interval $[0,t]$ (see Fig. \ref{Fig1}) let $\{\tau_1,\tau_2,\cdots,\tau_N \}$ denote the interval lengths
between successive zeros and $A(t)$
denote the length (or age) of the last {\em unfinished} excursion. The {\em extreme} observable we focus on is the length of the {\em longest}
excursion up to $t$
\begin{equation}
\label{def_lmax}
l_{{\rm max}}(t) = {\max} ( \tau_1, \tau_2, \cdots, \tau_N, A(t)) \;.
\end{equation} 

We  show here that  the average $\langle l_{\max}(t) \rangle$ is a quantity sensitive to 
 the non-Markovian character of fBm. In Ref. \cite{us_prl}, it was shown that $\langle l_{\max}(t) \rangle$ can be
conveniently computed using the exact relation
\begin{eqnarray}\label{rel_l_q}
d\langle l_{\max}(t)\rangle/dt = Q(t)\;,
\end{eqnarray}
where $Q(t)$ is the probability that the last unfinished excursion,
$A(t)$ in~Fig.~\ref{Fig1}, is the longest one
\begin{eqnarray} \label{def_Q}
Q(t) = {\rm Prob}[l_{\rm max}(t) = A(t)] \;.
\end{eqnarray}
It was then shown that for a renewal process characterized by a persistence exponent $\theta < 1$, one has the exact result \cite{us_prl}
\begin{eqnarray}\label{q_inf_renewal}
&&\lim_{t\to \infty} Q(t) = Q^R_{\infty} \nonumber \\
&&Q_\infty^R \equiv  Q^R_{\infty}(\theta) = \int_{0}^\infty \frac{dx}{ 1 + x^{\theta} e^x  \int_0^x \, dy \, y^{-\theta} e^{-y}} \,,
\end{eqnarray}
where the superscript 'R' refers to renewal process. In this paper, we
compute numerically $Q(t)$ (\ref{def_Q}) for fBm defined as in
Eq. (\ref{correl}) for different values of $0 < H < 1$. We show that,
in all these cases, $Q(t) \to Q_\infty$ for large time $t$, as predicted
in Ref. \cite{us_prl} for non-smooth processes with $0< \theta < 1$, which is 
the case for fBm with $0 < H < 1$. We then extract precisely the asymptotic value
$Q_\infty \equiv Q_\infty(H)$ :  any deviation from the value
$Q_\infty^R(\theta = 1-H)$ in Eq. (\ref{q_inf_renewal}) can thus be
identified as a signature of non-Markovian effects.  

{\em Numerical method.} For the purpose of numerical simulations we need to discretize the  fBm path into a set of
Gaussian numbers correlated through Eq. (\ref{correl}). Generating a sequence  ${x}=\left\lbrace 
x_1, ...,x_i, ... , x_{T} \right\rbrace $  of Gaussian numbers with prescribed correlations $\langle x_i x_j \rangle = C_{i,j}$
is a two step procedure: $i)$ it is first necessary to compute the matrix  $A$, the square root  of
the correlation $C$. $ii)$  Each discrete path is then given by ${x}=A {\xi}$, where  
${\mathbf \xi} = \left\lbrace \xi_1,\xi_2,\ldots,\xi_T \right\rbrace$ is an uncorrelated normally distributed set of random variables.  It is easy to check that paths obtained from this procedure have the required correlation matrix:
\begin{equation}
\langle x_i x_j \rangle = \sum_{{k_1,k_2}=1}^T  A_{i,k_1}    A_{k_2,j}  \langle  \xi_{k_1}\xi_{k_2} \rangle =A^2_{i,j}=C_{i,j}.
\end{equation}
Compared to standard  Brownian motion, building a fBm is numerically cumbersome.
The Brownian motion has a linear cost in $T$ and is easy to simulate paths of  size $T\sim 10^6$.  For fBm, the first step 
 involves the full diagonalization of matrix $C_{i,j}$ and  limitates to $T\sim 1000$ the size of the path. The second step is faster and the matrix-vector product needs $T^2$ operations. A better performance can be obtained for fBm thanks to the stationarity of  the incremental  correlation function. The increments $\delta_i=x_{i+1}-x_i$ are correlated according to a Toeplitz matrix. For Toeplitz matrices special numerical methods allow to build paths without going through the full diagonalization of $C$. Here we use  the Levinson algorithm which is not the fastest algorithm, but is exact for any value of $T$  (for a pratical implementation see \cite{dieker}). In this paper we show the results obtained for fBm of size $T=10000$.

\begin{figure}
\centerline{
\includegraphics[width=\columnwidth]{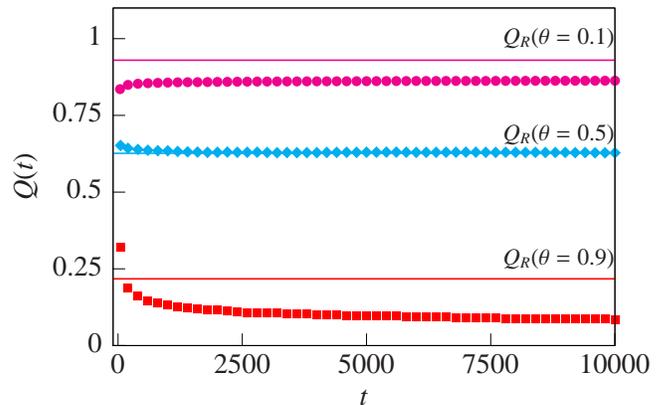}}
\caption{$Q(t)$ as a function of $t$ for $H = 0.1, 0.5$ and $H=0.9$. The straight lines correspond to the value of $Q_\infty^R(1-H)$ for a renewal process given in Eq. (\ref{q_inf_renewal}). This clearly illustrates that the fBm is not a renewal process.}
\label{figure_qinf}
\end{figure}

{\it Numerical results.} We now discuss our results for $Q(t)$ defined in Eq. (\ref{def_Q}), whic was computed by averaging over $10^6$ samples. In Fig. \ref{figure_qinf} we show a plot of $Q(t)$ as a function of $t$ for different values of $H = 0.1, 0.5$ and $H=0.9$. In all these cases our numerical data are consistent with an asymptotic behavior 
\begin{eqnarray}\label{asympt_behav}
\lim_{t \to \infty} Q(t) = Q_{\infty} \equiv Q_\infty(H) \;.
\end{eqnarray}
We also notice that this asymptotic value is approached from above for $H=0.1, 0.5$ and from below for $H=0.9$. 
In this same figure \ref{figure_qinf} we also plot, with dotted lines,
the value of $Q_\infty^R$ for a renewal process given in
Eq. (\ref{q_inf_renewal}) with $\theta = 1-H$.  These two values
$Q_\infty$ and $Q_{\infty}^R$ are clearly different as $H$ deviates
significantly from $1/2$. We have also checked that for all these
values of $H$, the persistence probability $p_0(t) \sim t^{-1+H}$
displays a well developed power law behavior for $t \geq 1000$ so that
a comparison with $Q_\infty^R(\theta=1-H)$ is meaningful. Therefore we
can conclude safely that $Q_\infty$ carries the signature of memory
effects of the fBm for $H \neq 1/2$.


\begin{figure}
\includegraphics[width=0.97\columnwidth]{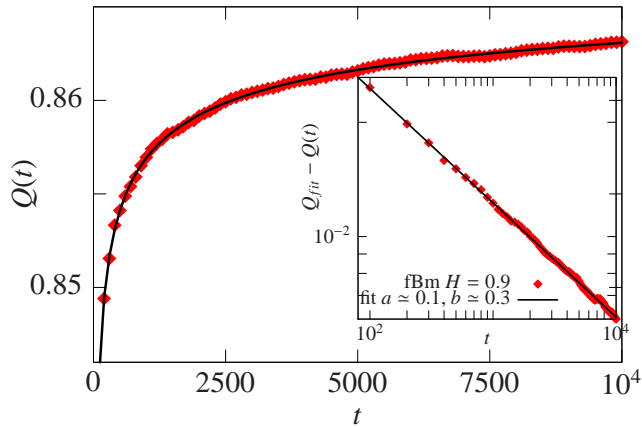}
\caption{Plot of $Q(t)$ as a function of $t$ on a linear-linear
  scale for $H=0.9$. The solid line indicates the fit as in
  Eq. (\ref{correc}) with $a(0.9) \simeq 0.1 $ and $b(0.9) \simeq
  0.3$. {\bf Inset} : Plot of $Q(t) - Q_\infty(H=0.9)$ (same data as in
  the main figure) as a function of
  $t$ in a log-log plot. The solid line corresponds to
  $a(0.9)t^{-b(H)}$ : this suggests a good quality of the fitting
  procedure in Eq. (\ref{correc}).} \label{figure_finitesize}  
\end{figure}

Although the curves for $Q(t)$ shown in Fig. \ref{figure_qinf}
 indicate an asymptotic behavior as in Eq. (\ref{asympt_behav}) with a
 value for $Q_\infty(H)$ different from $Q^R_\infty(1-H)$, a precise
 estimate of this asymptotic value $Q_\infty(H)$ requires more
 effort. The time dependence of $Q(t)$ is due to discretization of the
 paths and this can be understood by studying the case of Brownian motion
 (BM). It is well known that the density of zero crossings of BM is
 infinite : this means that if the BM crosses zero once, it will
 recross zero infinitely many times immediately after the first
 crossing. Therefore a proper definition of the excursions requires a
 regularization procedure. A convenient way to implement it, is to
 impose that the maximal distance from the origin during an excursion
 should be bigger than $x_0$, where $x_0$ plays the role of a spatial
 cutoff. To compute the finite time behavior of $Q(t)$ we recall that
 the probability $p_0(t,x_0)$ that a BM starting from $x_0$ at $t=0$ remains
 positive up to time $t$ (persistence probability) is given by 
\begin{equation}\label{asympt_p0} 
p_0(t,x_0) \equiv
 p_0\left(\frac{t}{x_0^2}\right)=\text{Erf}\left(\frac{x_0}{
 \sqrt{2t}}\right) = \sqrt{\frac{{2}x_0^2}{{\pi t}}} + {\cal
 O}\left(\frac{x_0^2}{t^{3/2}} \right) 
\end{equation}
Following the derivation of Ref. \cite{us_prl} we can compute the
Laplace transform $\hat{Q}(s)$ of $Q(t)$ in the limit $x_0^2 \ll t$ \footnote{This formula is valid up to terms of order ${\cal O}(x_0^2/t)$. The precise evaluation of these terms is however quite difficult and goes beyond the scope of this paper.} 
\begin{equation}\label{laplace_q}
\hat{Q}(s)=\frac{1}{s} \int_0^\infty d x \frac{p_0(x/s) e^{-x}}{p_0(x/s) e^{-x}+\int_0^x d y p_0(y/s) e^{-y}}.
\end{equation}
For small $s$, it was shown in Ref. \cite{us_prl} that $\hat{Q}(s)
\sim {Q_{\infty}^R(1/2)}/{s}$ where $Q_\infty^R(1/2) =0.626508...$
\cite{pitman_yor}. To understand the effects of the
discretization, one needs to compute the first correction to this
leading $1/s$ behavior when $s \to 0$. This has to be done carefully because
a naive expansion of the persistence probability $p_0(y/s)$ beyond the
leading order as suggested by Eq. (\ref{asympt_p0}) in the denominator
of Eq. (\ref{laplace_q}) yields a diverging integral
over $y$. Handling this singular behavior with care yields   
\begin{equation}
\hat{Q}(s)=\frac{Q_{\infty}^R(1/2)}{s} + \tilde a \frac{x_0}{\sqrt{s}} +
    {\cal O}(1) \;,
\end{equation}
where $\tilde a$ is given by
\begin{equation}
\tilde a = \int_0^\infty \frac{e^{-x} x^{1/2}}{\left(x^{-1/2}e^{-x}+
 \sqrt{\pi} {\rm Erf}{(\sqrt{x})}\right)^2}dx = 0.23970... \;.
\end{equation}
Going back to real time this yields finally
\begin{eqnarray}\label{correc_bm}
 Q(t) = Q_\infty^R(1/2) + \tilde a \sqrt{\frac{x_0^2}{t}} +
 {\cal O}(x_0^2/t)\;.
\end{eqnarray}

\begin{figure}
\centerline{
\includegraphics[width=1.05\columnwidth]{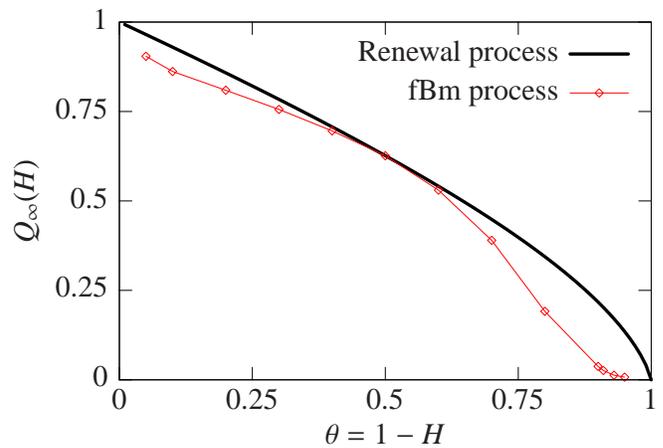}}
\caption{The triangles indicate the numerical estimate of $Q_\infty(\theta
  = 1-H)$, extracted from the fitting procedure in
  Eq. (\ref{correc}). For comparison, we have also plotted
  $Q_{\infty}^R(\theta = 1-H)$ for a renewal process, as given by
  Eq. (\ref{q_inf_renewal}). This plot clearly shows that, except for
  $H = 1/2$, the fBm is not a renewal process.}
\label{fig_comp}
\end{figure}

Motivated by this result for Brownian motion (\ref{correc_bm}), we
propose to describe the data for $Q(t)$ in Fig. (\ref{figure_qinf}) by
the following form
\begin{eqnarray}\label{correc}
Q(t) \sim Q_\infty(H) + a(H) t^{-b(H)} \;.
\end{eqnarray}
In particular, from Eq. (\ref{correc_bm}), one expects $b(1/2)
= 1/2$. We have checked that this form (\ref{correc}) describes very well our
data for $Q(t)$ for all the values of $0 < H < 1$ that we have
studied. In the inset of Fig. \ref{figure_finitesize}, we show a plot
of $Q(t) - Q_\infty(H=0.9)$, as a function of $t$ on a log-log
scale, while the main figure shows a plot of $Q(t)$ as a function of
$t$ on a linear-linear plot. This fitting procedure (\ref{correc}) 
hence provides a reliable way to estimate the asymptotic value
$Q_\infty(H)$. In Fig. \ref{fig_comp} we have plotted these values as a
function of $\theta = 1-H$. For comparison, we have also plotted the
values of $Q_\infty^R(\theta)$ for renewal process
(\ref{q_inf_renewal}) : these two curves are clearly different (except
for $H=1/2$ which corresponds to Brownian motion). 

Our numerical data indicate that the exponent $b(H)$ exhibits a
maximum for $H \sim 0.5$, where $b(1/2)= 1/2$. On the other hand, one finds that $b(H) \to 0$ for
$H \to 0$ and $H \to 1$ : therefore it 
becomes very difficult to extract a reliable value for $\theta$ close
to $0$ and $1$. For $H = 1$, the fBm is simply a linear function of
time $t$, $x(t) = \zeta t$ where $\zeta$ is Gaussian random variable of
unit variance. It is thus easy to see that $Q_\infty = 1$ in that
case. Although it is very difficult to extract a reliable value of
$Q_\infty(H)$ for $H > 0.95$, one expects that $Q_\infty(H) \to 1$,
smoothly, when $H \to 1$. Similarly, our data suggest that
$Q_\infty(H)$ vanishes smoothly as $H \to 0$. Finally, we notice that
$a(H)$ changes sign for $H \sim 0.7$ : it is positive for $H \gtrsim 0.7$
and negative for $H \lesssim 0.7$.

{\it Conclusion.} To conclude, we have presented a numerical
computation of the mean longest excursion $\langle l_{\rm max}(t)
\rangle$ for the fBm with Hurst index
$0 < H < 1$. We have shown that $\langle l_{\max}(t) \rangle \sim Q_\infty(H)t$ for large $t$
where $Q_\infty(H)$ is a new interesting feature of fBm. We have also demonstrated that this quantity is very sensitive to
temporal correlations characterizing this process. Therefore, at
variance with the recent claim of Ref. \cite{cakir_renewal}, our
numerical results clearly show that the zero crossings of fBm can not be described by a
renewal process. We point out that the quantity studied here is sensitive
to the full joint distribution of the time intervals between
crossings, while the numerical work presented in
Ref. \cite{cakir_renewal} only studied the correlation function
between two such intervals. Finally we hope that the non trivial dependence of
$Q_\infty(H)$ shown in Fig. \ref{fig_comp} will stimulate further
analytical progress on the study of fBm.

\begin{acknowledgments}
This work was supported by the France-Argentina MINCYT-ECOS A08E03. We thank
Joachim Krug for pointing out Ref. \cite{cakir_renewal}. We acknowledge C. Godr{\`e}che and
S.N. Majumdar for stimulating discussions at the earliest stage of this work. R.G.G. acknowledges support from CONICET and the hospitality at LPT and LPTMS in Orsay.  
\end{acknowledgments}

\end{document}